\documentclass[prd,onecolumn,nofootinbib]{revtex4}
\usepackage{bm,amsmath,amssymb,graphicx}

\begin{document}

\title{The Origin of the Born Rule from Spacetime Averaging}
\author{Nikodem Pop{\l}awski$^1$}
\altaffiliation{NPoplawski@newhaven.edu}
\author{Michael Del Grosso$^2$}
\altaffiliation{mdelg3@unh.newhaven.edu}
\affiliation{$^{1}$Department of Mathematics and Physics, College of Arts and Sciences}
\affiliation{$^{2}$Department of Chemical Engineering, Tagliatela College of Engineering, University of New Haven, 300 Boston Post Road, West Haven, CT 06516, USA}


\begin{abstract}
The Born rule postulates that the probability of measurement in quantum mechanics is related to the squared modulus of the wave function $\psi$.
We rearrange the equation for energy eigenfunctions to define the energy as the real part of $\hat{E}\psi/\psi$.
For an eigenstate, this definition gives a constant energy eigenvalue.
For a general wave function, the energy fluctuates in space and time.
We consider a particle in a one-dimensional square well potential in a superposition of two states and average the energy over space and time.
We show that, for most cases, such an energy expectation value differs by only a few percent from that calculated using the Born rule.
This difference is consistent with experimental tests of the expectation value and suggests that the Born rule may be an approximation of spacetime averaging.
\end{abstract}
\maketitle

In nonrelativistic quantum mechanics, three physical postulates derive a complete interpretation of the wave function \cite{Schiff}.
Firstly, each dynamical variable related to the motion of a particle is represented by an operator.
Each operator is associated with an eigenvalue equation.
The energy operator is $\hat{E}=i\hbar\,\partial/\partial t$ and the eigenvalue equation is
\begin{equation}
\hat{E}u_E=Eu_E.
\label{eigen}
\end{equation}
The functions $u_E$ that satisfy this equation are the energy eigenfunctions and the corresponding values of $E$ are the energy eigenvalues.
Secondly, the only possible result of a precise measurement of a dynamical variable (observable) is one of the eigenvalues.
Thirdly, the number of measurements resulting in an eigenvalue $E_n$ is proportional to the square of the modulus of the coefficient of $u_{En}$ in the expansion of the wave function $\psi$ into the eigenfunctions.
The last postulate represents the Born rule \cite{Born}.

Rearranging the equation for energy eigenfunctions (\ref{eigen}) gives $E=(\hat{E}u_E)/u_E$.
We propose that this formula can be used to define the energy for a general wave function that is not an energy eigenfunction:
\begin{equation}
E=\mbox{Re}\Bigl(\frac{\hat{E}\psi}{\psi}\Bigr).
\label{energy}
\end{equation}
For an eigenstate, this definition gives a constant energy eigenvalue that does not depend on the space and time coordinates.
Consequently, the second postulate is satisfied automatically.
For a general wave function, the energy is not constant and fluctuates in space and time.
During a measurement, the energy (beable) changes into one of the eigenvalues (observables) according to some unknown process referred to as the wave function collapse \cite{QM}.
Substituting $\psi=\exp(iS/\hbar)$, where $S$ is the action for a particle \cite{LL1}, to formula (\ref{energy}) gives a classical mechanics relation:
\[
E=-\frac{\partial S}{\partial t}.
\]
Because the energy operator is differential, equation (\ref{energy}) can also be written as $E=\mbox{Re}[\hat{E}(\ln\psi)]$.

We consider a linear superposition of two states with energies $E_1,E_2$ and real energy eigenfunctions $u_{E1}(x),u_{E2}(x)$:
\begin{equation}
\psi(x,t)=e^{-i\omega_1 t}f(x)+e^{-i\omega_2 t}g(x),
\label{wave}
\end{equation}
where $f(x)=c_1 u_{E1}(x)$, $g(x)=u_{E2}(x)$, $c_1,c_2$ are complex constants, which can be assumed real without loss of generality, $\omega_1=E_1/\hbar$, and $\omega_2=E_2/\hbar$.
Substituting the wave function (\ref{wave}) into the definition of the energy (\ref{energy}) gives
\begin{equation}
E=\mbox{Re}\Bigl(\frac{E_1 e^{-i\omega_1 t}f+E_2 e^{-i\omega_2 t}g}{e^{-i\omega_1 t}f+e^{-i\omega_2 t}g}\Bigr)=\frac{E_1 f^2+E_2 g^2+(E_1+E_2)fg\cos(\omega t)}{f^2+g^2+2fg\cos(\omega t)}=\frac{E_1+E_2}{2}+\frac{(E_1-E_2)(f^2-g^2)}{2(f^2+g^2)(1+a\cos(\omega t))},
\label{twostates}
\end{equation}
where $a=2fg/(f^2+g^2)$ and $\omega=\omega_2-\omega_1$.

We propose that the expectation value $\langle E\rangle$ for the energy is calculated by averaging the energy in equation (\ref{energy}) over space and time.
For a superposition of two states, the energy $E$ is given by equation (\ref{twostates}).
Its time average is equal to $\bar{E}=\int_0^T E dt/T$, where $T=2\pi/\omega$.
Using the integral $\int_0^T[1+a\cos(\omega t)]^{-1}dt=T/\sqrt{1-a^2}$ gives
\begin{equation}
\bar{E}=\frac{E_1+E_2}{2}+\frac{E_1-E_2}{2}\frac{f^2-g^2}{f^2+g^2}\Bigl|\frac{f^2+g^2}{f^2-g^2}\Bigr|=\frac{E_1+E_2}{2}+\frac{E_1-E_2}{2}\mbox{sgn}(f^2-g^2).
\label{time}
\end{equation}
In particular, if $E_1=0$ and $E_2=E_0$, then $\bar{E}=0$ for $f^2>g^2$, $\bar{E}=E_0$ for $g^2>f^2$, and $\bar{E}=E_0/2$ for $f^2=g^2$.
For these eigenvalues, the time average of the energy is equal to the eigenvalue corresponding to the eigenfunction which has a greater modulus in the superposition.
Averaging $\bar{E}$ over space gives $\langle E\rangle/E_0$ equal to the fraction of space where $g^2>f^2$.

To illustrate how the spacetime-average expectation value differs from that derived using the Born rule, we consider a particle with mass $m$ in a one-dimensional square well potential with length $L$.
The wave function $\psi(x,t)$ for this particle satisfies the Schr\"{o}dinger equation:
\[
i\hbar\frac{\partial\psi}{\partial t}=-\frac{\hbar^2}{2m}\frac{\partial^2 \psi}{\partial x^2},
\]
with the boundary conditions $\psi(0,t)=\psi(L,t)=0$ \cite{Schiff,QM}.
The solutions of this linear equation are stationary states, represented by $\psi(x,t)=\exp(-iEt/\hbar)u_E(x)$,
where $E=(n^2 \pi^2 \hbar^2)/(2mL^2)$, $u_E(x)=\sqrt{2/L}\sin(n\pi x/L)$ are real energy eigenfunctions, and $n$ is a positive integer.

Averaging $\bar{E}$ in equation (\ref{time}) over the length of the well gives the spacetime-average energy expectation value for a superposition of two eigenstates (\ref{wave}):
\begin{eqnarray}
& & \langle E\rangle_\textrm{DGP}=\frac{1}{L}\int_0^L\bar{E}dL=\frac{E_1+E_2}{2}+\frac{E_1-E_2}{2L}\int_0^L\mbox{sgn}\bigl(f^2(x)-g^2(x)\bigr)dx \nonumber \\
& & =\frac{1}{LT}\int_0^L\int_0^T\mbox{Re}\Bigl(\frac{E_1 e^{-i\omega_1 t}f+E_2 e^{-i\omega_2 t}g}{e^{-i\omega_1 t}f+e^{-i\omega_2 t}g}\Bigr)dt\,dx,
\label{space}
\end{eqnarray}
where
\[
f(x)=c_1\sqrt{\frac{2}{L}}\sin\Bigl(\frac{n_1\pi x}{L}\Big),\quad g(x)=c_2\sqrt{\frac{2}{L}}\sin\Bigl(\frac{n_1\pi x}{L}\Big).
\]
The quantity (\ref{space}) is finite, even though the energy (\ref{energy}) is undefined at points where $\psi=0$.
The Born expectation value for the particle is equal to
\[
\langle E\rangle_\textrm{B}=\frac{c_1^2 E_1+c_2^2 E_2}{c_1^2+c_2^2}.
\]
For $c_1=0$ or $E_1=E_2$, both expectation values give the same result: $\langle E\rangle_\textrm{DGP}=\langle E\rangle_\textrm{B}=E_2$, corresponding to a pure eigenstate.

The expectation values for a superposition of more eigenstates are
\[
\langle E\rangle_\textrm{B}=\frac{\sum_n c_n^2 E_n}{\sum_n c_n^2},\quad\langle E\rangle_\textrm{DGP}=\frac{1}{LT}\int_0^L\int_0^T\mbox{Re}\Bigl(\frac{\sum_n c_n E_n e^{-i\omega_n t}\sin(n\pi x/L)}{\sum_n c_n e^{-i\omega_n t}\sin(n\pi x/L)}\Bigr)dt\,dx,
\]
where $T$ is the least common multiple of the values $2\pi/\omega$ calculated for each pair of the angular frequencies $\omega_n=E_n/\hbar$.

Equation (\ref{space}) shows that the spacetime-average energy expectation value depends on the locations of the intersections of the squares of the sine functions $f(x)$ and $g(x)$, which determine the regions where one function has a greater modulus than the other.
This equation also shows that the spacetime-average energy expectation value depends on $n_1$ and $n_2$ only through their ratio $n_2/n_1$ because multiplying $n_1$ and $n_2$ by the same integer factor results in $f(x)$ and $g(x)$ oscillating in space faster by the same factor without changing the fraction of space where $g^2>f^2$.

\begin{figure}[h]
\includegraphics[width=2.5in]{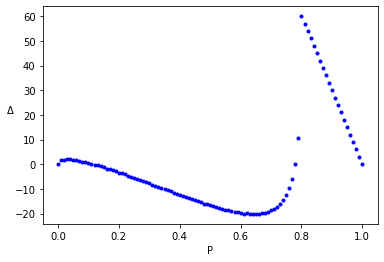}
\caption{$\Delta(P)$ for $n_1=1,\,n_2=2$.}
\label{1_2}
\end{figure}

\begin{figure}[h]
\includegraphics[width=2.5in]{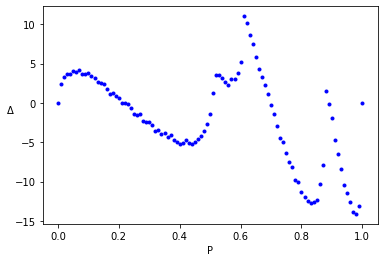}
\caption{$\Delta(P)$ for $n_1=3,\,n_2=8$.}
\label{3_8}
\end{figure}

\begin{figure}[h]
\includegraphics[width=2.5in]{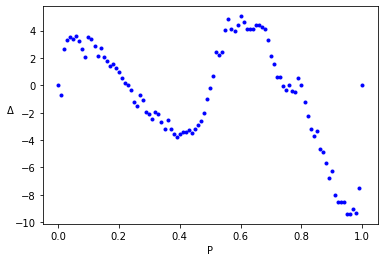}
\caption{$\Delta(P)$ for $n_1=13,\,n_2=25$.}
\label{13_25}
\end{figure}

\begin{figure}[h]
\includegraphics[width=2.5in]{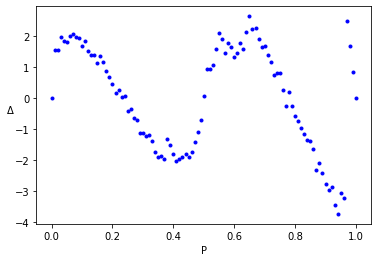}
\caption{$\Delta(P)$ for $n_1=17,\,n_2=23$.}
\label{17_23}
\end{figure}

\begin{figure}[h]
\includegraphics[width=2.5in]{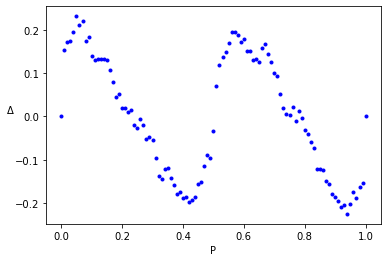}
\caption{$\Delta(P)$ for $n_1=42,\,n_2=43$.}
\label{42_43}
\end{figure}

We present several figures showing the relative difference between both expectation values for linear combinations of two normalized eigenstates with $c_1^2+c_2^2=1$ for various ratios $n_2/n_1$.
They show the relative difference,
\[
\Delta=\frac{\langle E\rangle_\textrm{B}-\langle E\rangle_\textrm{DGP}}{\langle E\rangle_\textrm{DGP}}\times100\,\%,
\]
as a function of $P=c_1^2$.
Figure \ref{1_2} shows $\Delta(P)$ for $n_1=1$ and $n_2=2$.
Figure \ref{3_8} shows $\Delta(P)$ for $n_1=3$ and $n_2=8$.
Figure \ref{13_25} shows $\Delta(P)$ for $n_1=13$ and $n_2=25$.
Figure \ref{17_23} shows $\Delta(P)$ for $n_1=17$ and $n_2=23$.
Figure \ref{42_43} shows $\Delta(P)$ for $n_1=42$ and $n_2=43$.

These results show that, for most cases, the energy expectation value derived from spacetime averaging differs by only a few percent from that calculated using the Born rule.
Such a difference is consistent with an experimental test of the expectation value of a physical variable describing a single particle and existing in two states: the polarization of a photon, which agrees with the Born rule within an uncertainty of 8 \% \cite{exp}.

These results also show that $\Delta$ generally decreases as the number of intersections of the squares of the functions $f(x)$ and $g(x)$ within $l$ increases, where $l$ is equal to $L$ divided by the greatest common divisor of $n_1$ and $n_2$.
The number of intersections generally increases as the minimum $\mbox{min}(n_1,n_2)$ in the simplified fraction $n_2/n_1$ increases.
Consequently, we expect that the relative difference between the Born and spacetime-average energy expectation values is smaller for linear combinations of more than two normalized eigenstates because their squared eigenfunctions have more intersections.
As the number of eigenstates in a linear combination tends to infinity, such a difference is expected to tend to zero.

These results suggest that the Born rule may be an approximation of spacetime averaging.
The third postulate in quantum mechanics, which assigns probabilities to quantum measurements, may therefore be redundant.
This conclusion agrees with approaches that attempt to deduce the mathematical structure of quantum measurements from the other quantum postulates \cite{red}.
One of those approaches uses the frequency operator \cite{freq}.
To complete quantum mechanics, one must discover a physical mechanism during a measurement that collapses the wave function into one of the eigenfunctions and turns the energy (\ref{energy}) into one of the eigenvalues.
Such a mechanism should illuminate the relation between spacetime averaging and ensemble averaging, which could be related through the quantum ergodic hypothesis \cite{ergo}, in calculating the expectation value.


\end{document}